\documentclass[twocolumn, superscriptaddress]{revtex4}
\pdfoutput=1
\usepackage{epsfig,graphicx}
\usepackage{amsmath,amssymb,amsfonts,amsthm}
\usepackage{bm} 
\usepackage{flafter}
\usepackage{color}
\usepackage{epstopdf}
\usepackage{latexsym}
\usepackage[colorlinks=true, citecolor=blue, urlcolor=blue ]{hyperref}
\usepackage{pxfonts,txfonts}
\usepackage{tgpagella}
\usepackage[T1]{fontenc}
\usepackage{ulem} 
\usepackage[utf8]{inputenc}

\DeclareMathOperator{\tr}{Tr}
\DeclareMathOperator{\ope}{\hat{O}}

\begin{document}

\title{
Using Random Boundary Conditions to simulate disordered quantum spin models in 2D-systems}
\author{A. Yuste}
\email{abel.yuste@uab.cat}
\affiliation{Departament de F\'{\i}sica, 
Universitat Aut\`{o}noma de Barcelona, 08193, Bellaterra, Spain.}
\author{M. Moreno-Cardoner}
\affiliation{Departament de F\'{\i}sica, 
Universitat Aut\`{o}noma de Barcelona, 08193, Bellaterra, Spain.}
\author{A. Sanpera} 
\affiliation{Departament de F\'{\i}sica, 
Universitat Aut\`{o}noma de Barcelona, 08193, Bellaterra, Spain.}
\affiliation{ICREA, Instituci\'o Catalana de Recerca i Estudis Avan\c cats, Pg. Llu\'is Companys 23, 08010 Barcelona, Spain.}

\begin{abstract}
Disordered quantum antiferromagnets in two-dimensional compounds have been a focus of interest in the last years due to their exotic properties. However, with very few exceptions,
the ground states of the corresponding Hamiltonians are notoriously difficult to simulate making their characterization and detection very elusive, both, theoretically and experimentally.
Here we propose a method to signal quantum disordered antiferromagnets by doing exact diagonalization in small lattices using random boundary conditions and averaging the observables of interest over the different disorder realizations. We apply our method to study a Heisenberg spin-1/2 model in an anisotropic triangular  lattice. In this model, the competition between frustration and quantum fluctuations might lead to some spin liquid phases as predicted from different methods ranging from spin wave mean field theory to 2D-DMRG or PEPS.  Our method accurately reproduces the ordered phases expected of the model and signals disordered phases by the presence of a large number of quasi degenerate ground states together with the absence of a local order parameter. The method presents a weak dependence on finite size effects. 

\end{abstract}
\maketitle

\section{Introduction}
\label{sec:Intro}
The characterization of disordered quantum antiferromagnets (AF) is currently one of the open challenges in modern condensed matter \cite{Lhuillier2012}. Ground states of ordered phases in quantum spin systems manifest themselves by long-range order (LRO) accompanied by the presence of local order parameters. The situation is drastically different when disorder arises due to quantum fluctuations and frustration, i.e., the impossibility to  simultaneously  minimize all local energy constrains.
Such disordered spin systems are expected to lack LRO and do not have local order parameters associated to them. Their presence can, in some cases, be confirmed by the topological entanglement entropy, a sub-leading term in the entanglement entropy which is invariant with the size of the plaquette \cite{Kitaev2006,Levin2006,Isakov2011}. However, determining the topological entanglement requires the precise knowledge of the ground state wave function which is often impossible due to the non-integrability of most AF frustrated models.
The importance of these phases, often dubbed topological, is both of fundamental and practical importance. They are at the forefront of present knowledge of strongly correlated systems and possess several features that make them very appealing for possible technological applications. They also lead to a reach variety of exotica phenomena as for instance fractional excitations and non-Abelian statistics \cite{Wen2016}.\\
Spin liquid (SL) phases are quantum disordered nonmagnetic phases that do not spontaneously break the spin rotation and discrete translational symmetry of the spin Hamiltonian \cite{Balents2010}. Recently, they have been proven to be the ground state of some Hamiltonians \cite{Kitaev2006a, Moessner2001} and SL candidates have been experimentally discovered in a vast range of materials \cite{Yamashita2010, Yamashita2008, DeVries2009, Shimizu2003, Mendels2007, Kurosaki2005, Itou2008, Helton2007, Han2012, Fak2012,Coldea2001}. Its existence seems to be intimately related to geometrical frustrated systems. A prototype of frustrated spin models are antiferromagnetic Heisenberg Hamiltonians in the spatially anisotropic triangular lattice (SATL), where the anisotropy is due to the different spin couplings along the lattice directions. Recently, first attempts to understand such systems have been realized with ultracold bosonic atoms in optical lattices \cite{Struck2011}. Despite the apparent simplicity of the model, there is presently a clear disagreement in the phase diagram of the system. In particular, in the existence, extension and nature of the disordered phases. It has been conjectured that for such a model a quantum SL phase appears between commensurate and incommensurate order \cite{Hauke2013a}. Such claim is controversial in the present literature. The AF Heisenberg model in the SATL has been theoretically approached with different techniques. These include, among others, mean field methods such as modified spin-wave theory (MSWT) \cite{Hauke2010,Hauke2013a,Celi2016} or cluster mean-field approach \cite{Yamamoto2014, Moreno-Cardoner2014, Moreno-Cardoner2014a}; numerically variational methods like 2D-DMRG in a cylinder \cite{Weng2006,Weichselbaum2011,Hu2015}, projected entangled pair states (PEPS)  \cite{Schmied2008}, Variational Monte-Carlo \cite{Tocchio2009,Tocchio2013}. Exact diagonalization (ED)  in small plaquettes (see for instance \cite{Leung1993, Bernu1994,Weng2006,Heidarian2009, Hauke2010,Thesberg2014}) have also been usually used to constrast results with the much more sophisticated techniques mentioned above.\\
Here, we present a different way to approach generic disordered quantum spin systems by using random boundary conditions in an otherwise ED method.  As it is customary in the treatment of disordered systems, for each realization of the disorder, i.e., for each set of random boundary conditions, we calculate the observables of interest and perform at the end an average. When performing ED, one of the observables of interest that can be easily computed is the spin structure factor, $S(\vec k)$, over the first Brillouin zone which is also a very relevant measurement in the experiments. In the ordered phases, $S(\vec k)$ displays the relative orientation of the spins in the different lattice sites and its maxima straightforwardly translate into the ordered pattern that the spins acquire in the lattice. In the disordered phases, it is reasonable to expect that $S(\vec k)$ will blur the well defined peaks associated to ordered patterns substantially broadening  the maxima over the first Brillouin zone. Our approach, therefore, has a twofold purpose. First, to get rid off of the rigidity imposed by periodic or open boundary conditions which forces the system to order depending on the size and geometry of the plaquette. Second, to allow the possibility of a large degeneracy of ground states (each corresponding to a disorder realization) which, in turn, can translate into a broader $S(\vec k)$ in the Brillouin zone obtained from the average of the different realizations.
Our method is inspired by the work of Santos et al. \cite{Mendes-Santos2014} which used twisted random boundary conditions to study a Fermi-Hubbard model in a two-dimensional optical lattice. Their results for small lattices were in very good agreement with the ones obtained by Quantum Monte Carlo with much larger lattices and did not present the sign problem inherent in this technique. Here we show that ED with random boundary conditions for really small clusters consisting on only N=12, 16 or 24 spins, provides relatively independent cluster size results whereas the size and geometry of the cluster critically influence the OBC or PBC results.
We focus here in the AF spin-1/2 XY model in the SATL, although the method we present is completely generic and can be adapted to study any other quantum spin model.  The XY-SATL model has been recently addressed in \cite{Thesberg2014} using twisted boundary conditions but in a different spirit since the set of twisted boundaries has been used to select as the ground state the one that minimizes the energy (or the first excited state). Such a choice still depends strongly on the geometry and size of the plaquette and we will show that it might be insufficient to study the disordered phases.\\
Before proceeding further with the details of our study, we briefly outline our major results here. 
We analyze the quantum phase diagram of the XY model in the SATL using ED with random boundary conditions denoted generically by $\{\varphi_i\}$. For each boundary configuration we diagonalize the Hamiltonian and find the eigenstates and eigenvalues. Firstly, in the ordered phases of the model we find that there is, in general, a single random configuration that leads to the lowest ground state energy, $E_{0,min}$. Any other random configuration, $\{\varphi_l\}$, whose ground state energy $E_{0,l}$ is very close to $E_{0,min}$ has a large fidelity with the latter (overlap) $F_{min,l}=|\langle\Psi_{0,min}|\Psi_{0,l}\rangle |^{2}\simeq 1$. The expectation value of any observable, $\ope$, obtained from such close states fulfils that 
$\tr(\ope|\Psi_{0,l}\rangle\langle\Psi_{0,l}|)\sim \tr(\ope|\Psi_{0,min}\rangle\langle\Psi_{0,min}|)$. We also find in some ordered phases configurations whose ground state energies $E_{0,l}\simeq E_{0,min}$ and $F_{min,l}=0$. We observe that such configurations correspond to ground states whose spins are locally rotated but compatible with the given order, as it happens with the different chirality of the spiral phase. Secondly, for some values of the lattice anisotropy, we observe that there are {\it{many}} random boundary configurations, $\{\varphi_s\}$, whose ground state energies $E_{0,s}$ are quasi degenerate with the configuration leading to lowest energy $E_{0,min}$, but whose fidelity with the latter can take arbitrary values, i.e., $F_{min,s}\in (0,1)$.  Such energetically very close configurations, can have very different expectation values of the same observable. The values of the lattice anisotropy for which such effects are present are in very close agreement with the predicted values for quantum SL using PEPS \cite{Schmied2008}. Thirdly, the above features arise independently of the size/geometry of the cluster used to perform ED although finite size effects are present.\\
The manuscript is organized as follows. In Section II, we introduce the AF spin-1/2 XY model in the SATL and describe the-state-of-the-art concerning its phase diagram. In Section III, we introduce our approach and explain in which way random boundaries are imposed. In Section IV, we present and discuss our results and finally in Section V we conclude.

\section{AF Spin-1/2 XY model in the SATL}
\label{sec:model}
\begin{figure}
\centering
\includegraphics[width=\columnwidth]{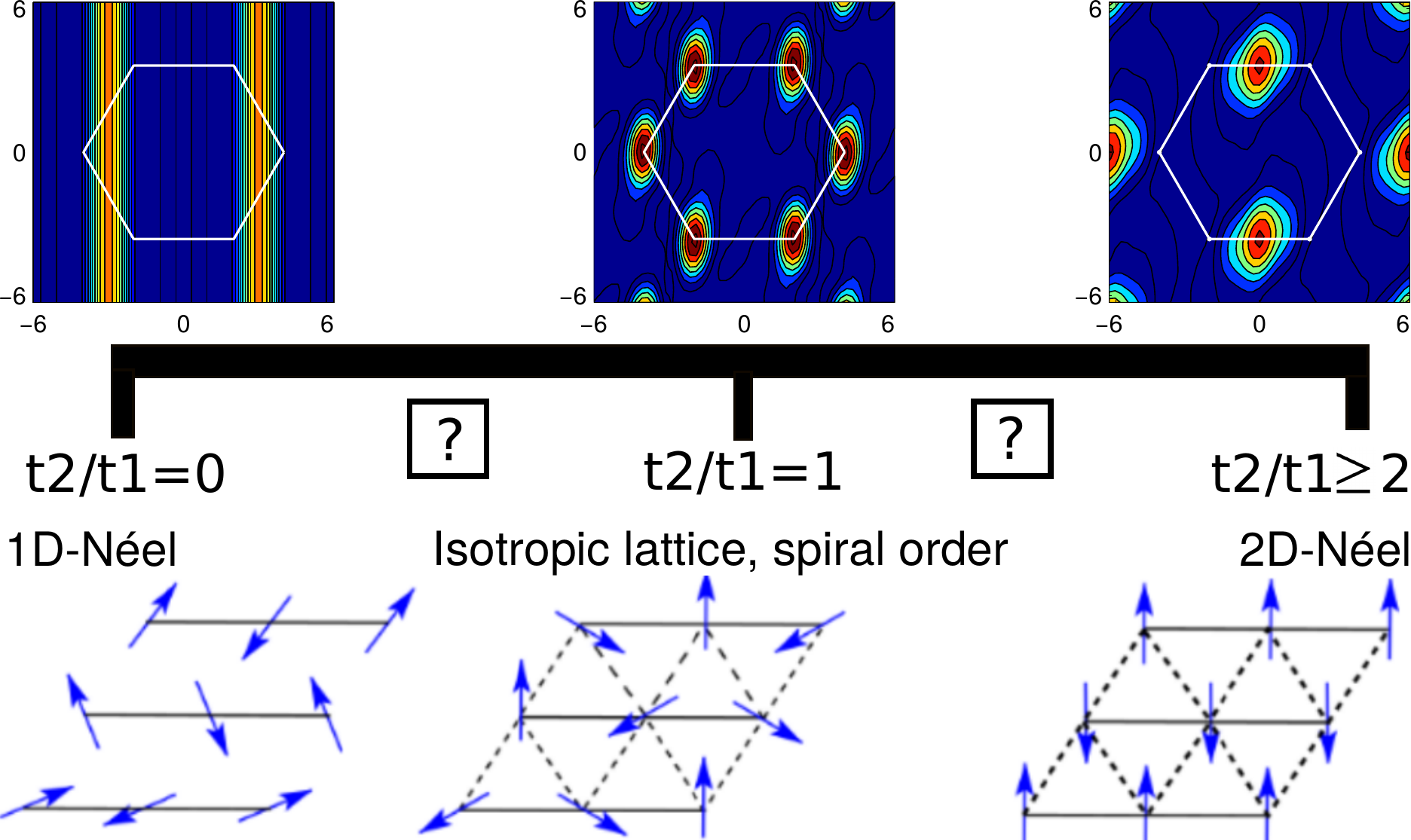}
\caption{Schematic phase diagram of the AF XY model in the SATL as a function of $t_2/t_1$. On the top, the spin structure factor, $S(\vec{k})$, for the three known ordered phases of the system: N\'eel order in the 1-D uncoupled chains ($t_2/t_1=0$), spiral order in the isotropic lattice ($t_2/t_1=1$) and N\'eel order between the different 1D chains for $t_2\ge 2t_1$. Bottom, schematic drawing of the spin directions in the above mentioned phases. The $?$ symbols mean that there is no consensus on the nature of these phases.}
\label{fig:phase_diagram}
\end{figure}

The Bose-Hubbard Hamiltonian in the triangular lattice reads
\begin{align}
\label{eq:Ham_BH}
\hat{H}_{BH}=\sum_{<i,j>}(t_{ij}\hat{b}^\dagger_i\hat{b}_j+h.c.)+\dfrac{U}{2}\sum_i\hat{n}_i(\hat{n}_i-1),
\end{align} 
where $\hat{b}^\dagger_i$($\hat{b}_i$) creates (annihilates) a boson on site $i$, $\hat{n}_i=\hat{b}^\dagger_i\hat{b}_i$ is the boson number operator, $t_{ij}$ is the tunneling parameter on the link $(i,j)$ and $U$ is the on-site repulsive interaction. The anisotropy of the model is given by $t_{ij}$ which translates into two different tunneling parameters: $t_1$ corresponding to tunneling along the horizontal link and $t_2$ for the diagonal ones, as indicated in Fig.\ref{fig:phase_diagram}. In the limit of hard-core bosons ($U\rightarrow \infty$), the model can be mapped onto a spin model using the Holstein-Primakov transformation, 
\begin{align*}
\hat b_{\alpha}^\dagger\longrightarrow S_{\alpha}^+=S_{\alpha}^x+iS_{\alpha}^y\\
\hat b_{\alpha}\longrightarrow S_{\alpha}^-=S_{\alpha}^x-iS_{\alpha}^y
\end{align*}
which maps creation and annihilation operators onto spin operators $S_{\alpha}$ and the Bose-Hubbard Hamiltonian becomes the spin-1/2 XY model,
\begin{align}
\label{XYHam}
\hat{H}_s=\sum_{<i,j>}t_{ij}(\hat{S}^x_i\hat{S}^x_j+\hat{S}^y_i\hat{S}^y_j).
\end{align}
Although both representations [\ref{eq:Ham_BH},\ref{XYHam}] are equivalent in the hard-core limit, in what it follows we diagonalize directly [\ref{eq:Ham_BH}]. In this limit, the second term in equation [\ref{eq:Ham_BH}] vanishes and the lattice filling factor is $\langle n_{i} \rangle =1/2$. To extract the ordering of the different phases it is standard to analyse two body correlations in momentum space,
\begin{align}
n(\vec k)={1\over N }\sum_{i\neq j}e^{-i\vec{k}(\vec{r}_i-\vec{r}_j)}\langle \hat{b}^\dagger_i\hat{b}_j +h.c\rangle,
\end{align}
which in the experiments with ultracold gases is obtained by means of the time-of-flight technique. This quantity straightforwardly maps onto the static structure factor for spin-1/2,
\begin{align}
n(\vec k)\rightarrow S(\vec k)={1\over N }\sum_{i\neq j}e^{-i\vec{k}(\vec{r}_i-\vec{r}_j)}\langle \hat{S}^{x}_i\hat{S}^{x}_j+\hat{S}^{y}_i\hat{S}^{y}_j\rangle,
\label{momentum}
\end{align}
where the sum extends to all the lattice sites of the cluster and the expectation value is taken over the ground state of the system. From $S(\vec k)$ it can be extracted both, the ordering vector $\vec{Q}=(Q_x,Q_y)$ which corresponds to the maxima of $S(\vec{k})$ and indicates classical order, and an order parameter, $M=\sqrt{S(\vec{Q})/N}$, which signals LRO in ordered states. Before proceeding further it is instructive to review the classical phase diagram which is obtained by replacing at each lattice site the spin operator for a classical rotor $ \vec{S}_{i}=S (Q_{x}^{clas} x_i, Q_{y}^{clas} y_i)$, up to a global phase factor. The classical ordering vector $\vec{Q}^{clas}$ lies in the XY plane and corresponds to the configurations that minimize the energy given by the Hamiltonian when the spin operators are replaced by spin vectors, $H=\sum t_{ij}\vec{S_i}\vec{S_j}$.  In spin wave theory,  such classical ordering is the reference state to which quantum corrections are added and then the energy is minimized self-consistently.
Minimization of the energy leads (at $Q_y=0$) to the following conditions:
\begin{eqnarray}
Q_{x}^{clas}&=& \pm\pi \;\; \mbox{\rm{for $t_2$}}=0 \cr
Q_{x}^{clas}&=&\pm2\arccos{(-\dfrac{t_2}{2t_1})} \;\; \mbox{\rm{for}}\;\;  0\leq t_{2}/t_{1}\leq 2\cr
Q_{x}^{clas}&=&\pm 2\pi  \;\; \mbox{\rm{for}}\;\;  t_{2}/t_{1}> 2.
\label{clasphasediagram}
\end{eqnarray}
The classical phase diagram of the model is sketched in Fig.\ref{fig:phase_diagram}, where the classical spin configurations together with their corresponding $S(\vec k)$ are displayed. Notice that at $t_2=0$, the system reduces to AF uncoupled 1D chains that order classically in the N\'eel configuration. For such order, $S(\vec k)$ has a maximum along $Q_{x}^{clas}= \pm \pi$ and is completely uncorrelated (disordered) along the $y$-direction.   At the isotropic point, $t_{2}/t_{1}=1$, the system has spiral order (N\'eel 120$^{\circ}$).  The maxima of $S(\vec k)$ are given now at the vertex of the hexagon $(Q_{x}^{clas},Q_{y}^{clas})=\pm \vec{b}_1; \pm \vec{b}_{2};(b_{2x},-b_{2y});(-b_{2x},b_{2y})$, provided by the triangular lattice. Here, $\vec{b}_1=(4\pi/3,0)$ and $\vec{b}_2=2\pi(1/3,1/\sqrt{3})$ are the inverse vectors of the lattice while the direct ones are given by $\vec{a}_1=(1,0)$ and $\vec{a}_2=(1/2,\sqrt{3}/2)$ with unit lattice constant. Finally, for $t_{2}\ge 2 t_{1}$, classically the AF order is given by a N\'eel  configuration along the diagonal chains, while horizontal chains display antiferromagnetic order, as schematically represented in Fig.\ref{fig:phase_diagram}. In this case, the triangular lattice becomes effectively a rombic one and $S(\vec k)$ displays the square order with maxima at $\vec{Q}^{clas}=(\pm 2\pi,0), (0,\pm 2\pi/\sqrt{3})$.\\ We have also depicted in Fig.\ref{fig:phase_diagram} the regions where there is presently not a clear consensus on the nature of the quantum phases. 
Lately, it has been pointed out the possibility that quantum SL phases appear in the transition from commensurate to incommensurate (i.e., the spins order with a period which is irrationally related to the lattice space) order. No direct evidence of such phases and their extension has already been unambiguously provided. 
Numerical results for the XY model using PEPS \cite{Schmied2008} supports such claim. However, MSWT \cite{Hauke2010} only finds signatures of the SL phase between the spiral and the 2D N\'eel phases. Recently, calculations for the XY model using ED with twisted boundary conditions \cite{Thesberg2014} and 2D DMRG calculations for the Heisenberg model in the SATL  \cite{Weichselbaum2011}  seem compatible with yet another ordering, the collinear antiferromagnetic (CAF) one.

\section{Method: ED with random boundary conditions}
\label{section:RBC}
Exact diagonalization is normally performed either on a cluster without considering links in the boundaries, the so-called open boundary conditions (OBC), or closing the plaquette with periodic conditions (PBC). For small clusters as the ones we are usually restricted for computational reasons, these boundary conditions have a strong influence on the results. That is so because both, the cluster geometry plus the boundary conditions act as a rigid  \textit{box} and the results obtained are conditioned by these two factors. Ideally, one should increase the size of the cluster used for ED until border effects become negligible, but this is usually not possible due to the fact that the corresponding Hilbert space grows exponentially imposing severe restrictions on computational resources. A way we believe can substantially mitigate the effect of the boundaries and the geometry is to simulate the disorder by imposing random boundary conditions. 
We define a set of random phases $\{\varphi_{ij}\}$ corresponding to complex tunneling elements hopping in/out from the cluster $t_{i,j}\rightarrow t_{i,j}e^{i\varphi_{ij}}$, $t_{j,i}\rightarrow t_{i,j}e^{-i\varphi_{ij}}$ if $(i,j)$ are links of the periodic boundary conditions. In the language of the Bose-Hubbard Hamiltonian this corresponds to hopping in an out of the cluster with a phase, in spin language this is equivalent to a spin twisting.

\begin{figure}
\centering
\includegraphics[width=\columnwidth]{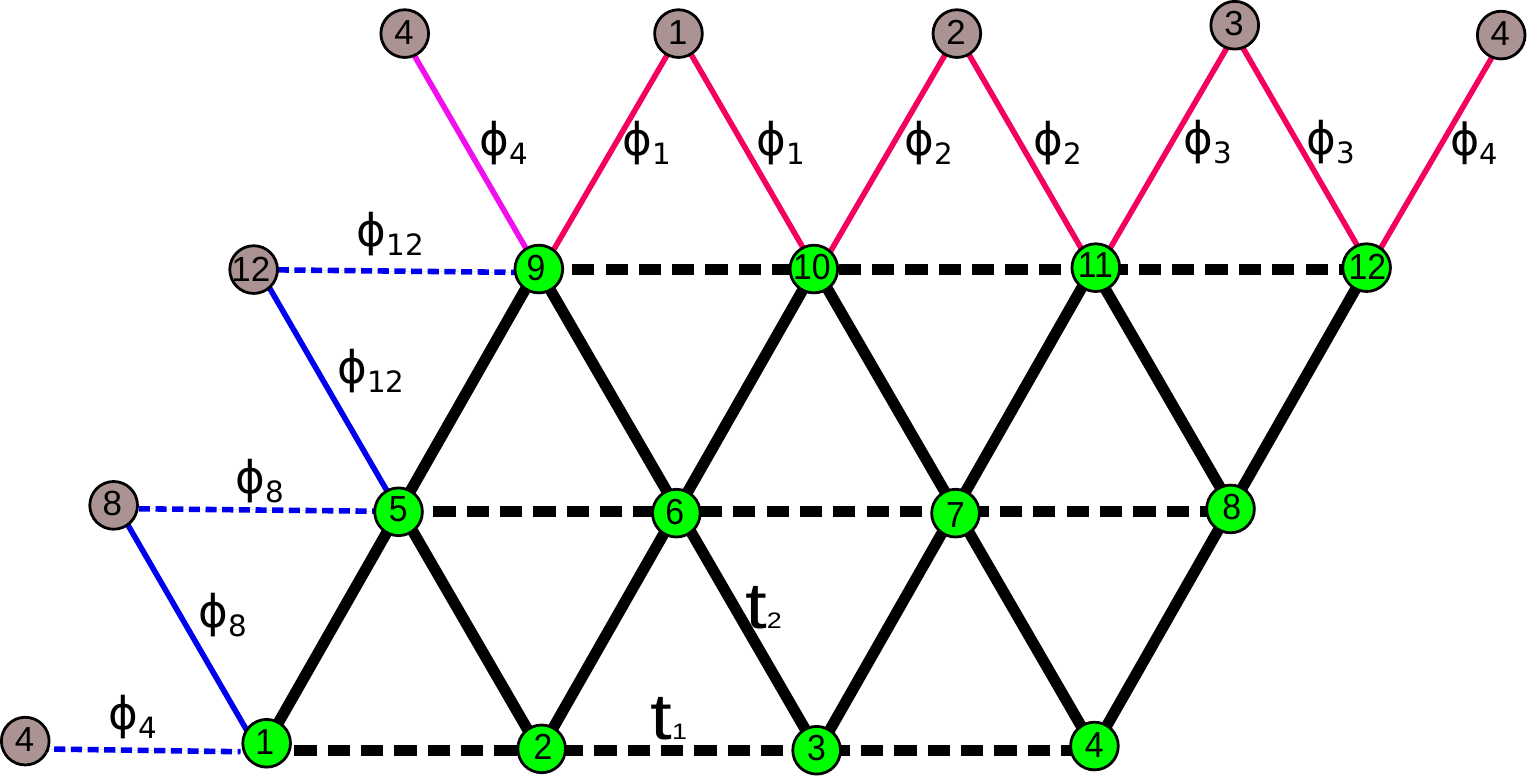}
\caption{Schematic representation of random boundary conditions (RBC and RRBC) in a 4x3  triangular lattice cluster. For the RBC, we assign two random phases at the boundary spins, $\phi_1$ for the blue links and $\phi_2$ for the red ones. Such a choice corresponds to two different twists on the lattice: one along the horizontal axis and the other one along the diagonal ones. At the corner of the lattice, the pink link acquires a phase $\phi_3=\phi_1+\phi_2$. 
For the RRBC, all the depicted tunneling in links from site $i$ acquire the same random phase $\phi_i$. The rest of the boundary links which are not shown are defined so to keep the Hermiticity of the Hamiltonian.}
\label{fig:lattice}
\end{figure}

We have used two criteria to define the random boundary conditions. 
The first one, denoted by RBC, has been used previously in \cite{Mendes-Santos2014, Thesberg2014}. In RBC, for each realization $i$ two different random phases are defined, $(\phi_1,\phi_2)_i$,  as sketched in Fig.\ref{fig:lattice}. The phase $\phi_1$ (blue lines) corresponds to tunneling out (in) of the cluster through the leftwards boundary links, $t_{1;2}e^{\pm i\phi_1}$, while the tunneling  associated to the upwards boundary links reads $t_2e^{\pm i\phi_2}$ (red lines). The link in the corner, since it can be interpreted as both a leftward and an upward, acquires a phase, $\phi_{3}=\phi_{1}+\phi_{2}$ (pink line). The rest of the boundary links are defined so to keep the Hamiltonian Hermitian. Such set of boundary conditions can be interpreted as a twist of the lattice along the directions determined by the direct vector of the lattice. \\
We have also considered a less restrictive configuration, denoted by RRBC and also schematically shown in Fig.\ref{fig:lattice} with the $\phi_i$ symbols. This boundary conditions cannot be interpreted as a twist of the lattice any more. However, they provide a larger flexibility on searching for disordered quantum spin systems.\\
With the above constrains, we have performed ED on rectangular clusters of  $N=L \times W$=12, 16 and 24 sites, where $L$ corresponds to the size of the chain and $W$ the number of chains in the plaquette. Diagonalization is done by keeping only the sector $S_{z}=0$ where the ground state lies. 
We generate a set of  random values, $\{\varphi_{k}\}_i$, and for each of them calculate the ground state energy $E_{0,i}$, the static spin structure factor $S_i(\vec{k})$, the ordering vector, $\vec{Q}_i$ and the order parameter $M_i=\sqrt{S_i(\vec{Q}_i)}/N$. All of them obviously depend on $\{\varphi_k\}_i$ and on the size and geometry of the plaquette. Finally, we perform the average of the quantities of interest over the disorder, which we denote with $\langle ...\rangle_d$. For each realization of RRBC, the set of random phases needed is sensibly larger than for RBC. For computational reasons we keep the number of realizations equal  to 200 hundred in both cases. Therefore, the results obtained from RRBC are less accurate. Nevertheless, as we shall see, our averaged results remain quite similar. 
\section{Results and Discussion}
\label{sec:results}
\begin{figure*}
\includegraphics[width=1.3\columnwidth ]{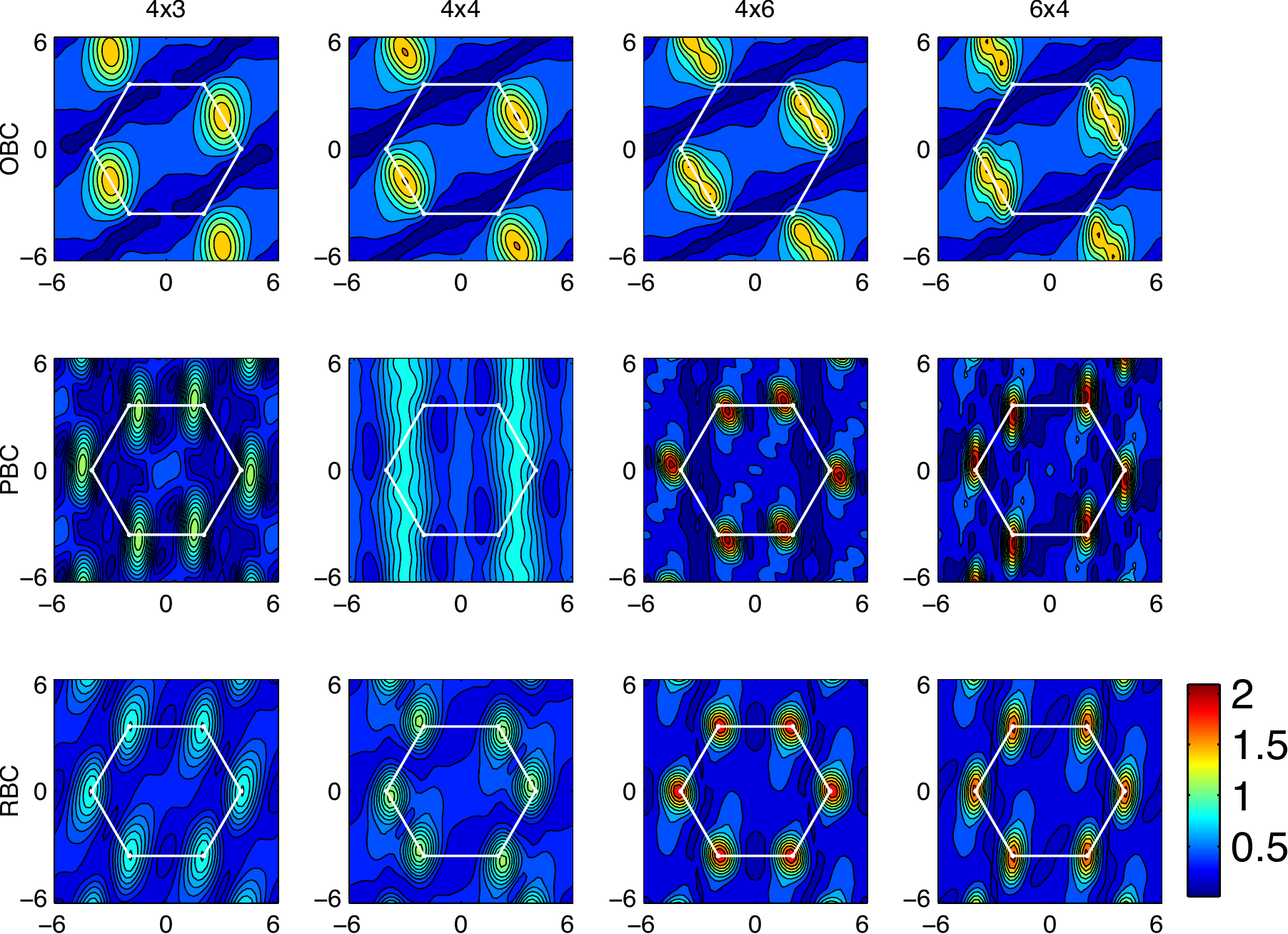}
\caption{Static spin structure factor $S(\vec{k})$ obtained from ED for different cluster geometries at the isotropic point, $t_2/t_1=1$ . Upper panel: Open Boundary Conditions. Middle panel: Periodic Boundary Conditions. Lower panel: Random Boundary Conditions. For the RBC, the averaged value over all the disorder, $\langle S(\vec{k})\rangle_d$, is depicted. The limit of the first Brillouin zone is shown with a white line.}
\label{fig:SK}
\end{figure*}

For simplicity, unless stated otherwise in what it follows we refer to RBC method. 
First thing which is worth noticing is the relative independence of the results on the geometry of the cluster. In Figure \ref{fig:SK}, we show the averaged static spin structure factor, $\langle S(\vec{k})\rangle_d$, at the isotropic point, $t_2/t_1=1$, for a cluster of 4x3, 4x4 and 6x4 sites and compare our results with the ones obtained imposing either OBC and PBC. On the one hand, the rigidity of OBC reflects into a wrong  CAF, whereas forces PBC to dramatically fail for the 4x4 lattice since such geometry suppresses spiral order. On the other hand, RBC is the only case which gives the correct order for all geometries, signalled by maxima at the corners of the first Brillouin zone.

More interesting is to look into the conjectured disordered quantum phases.
Notice that when sampling with random boundary conditions there can be configurations, $i$, whose corresponding ground state energy, $E_{0,i}$, substantially differs from the lowest achieved within the configurations, $E_{0,min}$. We define a renormalized energy, $\epsilon_i=\dfrac{E_{0,i}-E_{0,min}}{E_{0,min}}\cdot 100$, and consider just those configurations with $\epsilon_i<1$, i.e., with an energy not  $1\%$ larger than $E_{0,min}$.
As a first proof of concept, we plot the number of configurations, $N_c$, which lie in such interval as a function of $t_2/t_{1}$. Our method shows that, independently on the geometry of the lattice there are two regions around $t_{2}/t_{1}\sim 0.6$ and $\sim 1.5$ where there exist many configurations whose energy are close to the minimal one. These regions coincide with the conjectured SL phases predicted in the literature. For all other regions of the phase diagram --corresponding to the ordered phases-- the number of compatible configurations decreases keeping a flat structure. It is also interesting to show that, in agreement with all previous calculations, the ordered 2D-N\'eel order seems to appear already for values of $t_{2}/t_{1}\ge 1.7$ and stabilizes before its classical value,  $t_{2}/t_{1}\ge 2$, due to quantum fluctuations. From now on, all averaged quantities over the disorder, $\langle...\rangle_d$, will be considering just the $N_c$ configurations which fulfill $\epsilon_i<1$ for each value of $t_{2}/t_{1}$.\\
\begin{figure}
\includegraphics[width=0.8\columnwidth]{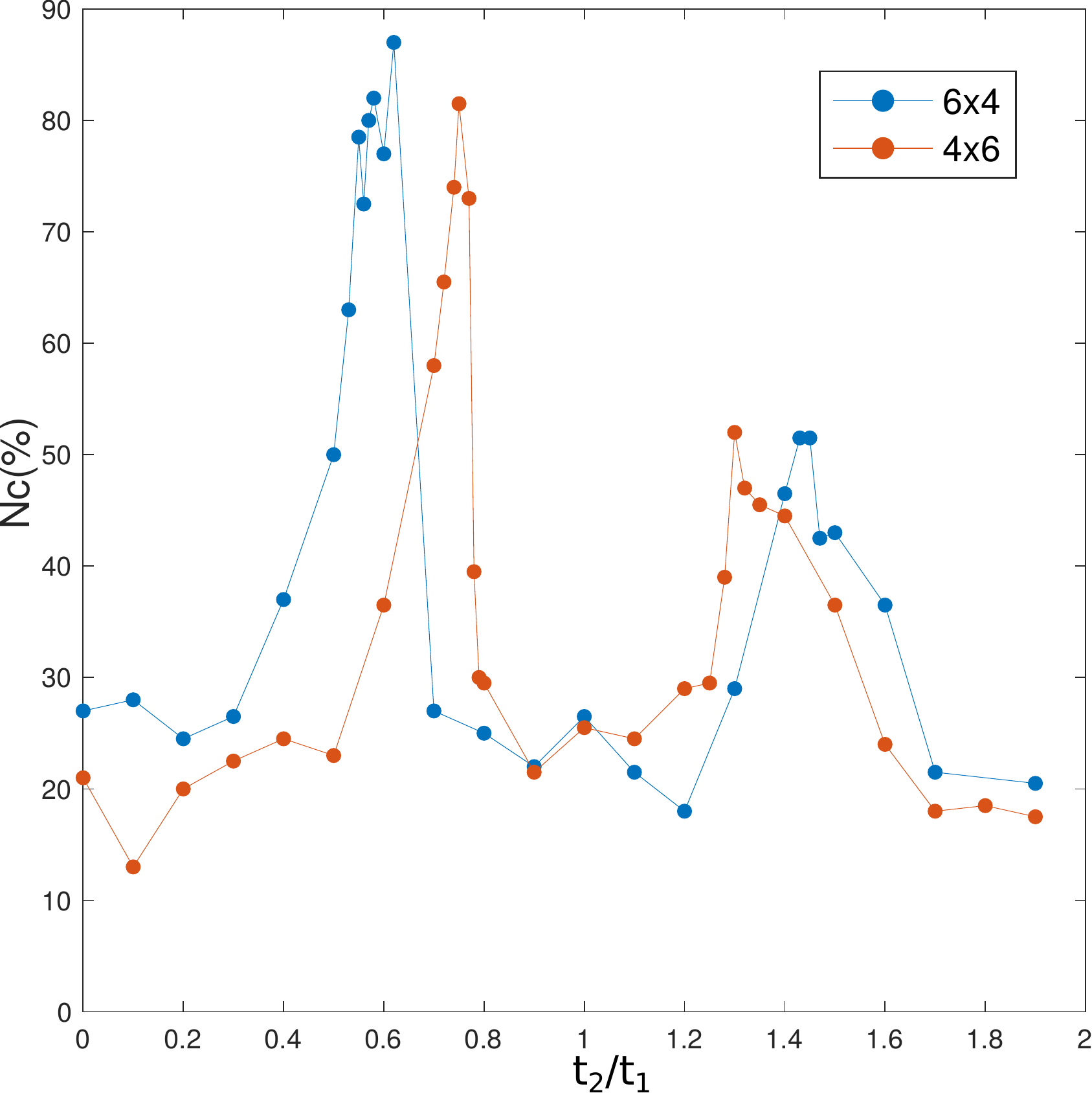}
\caption{Number of configurations, $N_c(\%)$, whose normalized ground state energy fulfills $\epsilon_i<1$ (see text) for a fixed value of the anisotropy $t_{2}/t_{1}$ using RBC. The two maxima signal the predicted quantum spin liquid phases.}
\label{fig:energy}
\end{figure}
\begin{figure*}
\includegraphics[width=2\columnwidth]{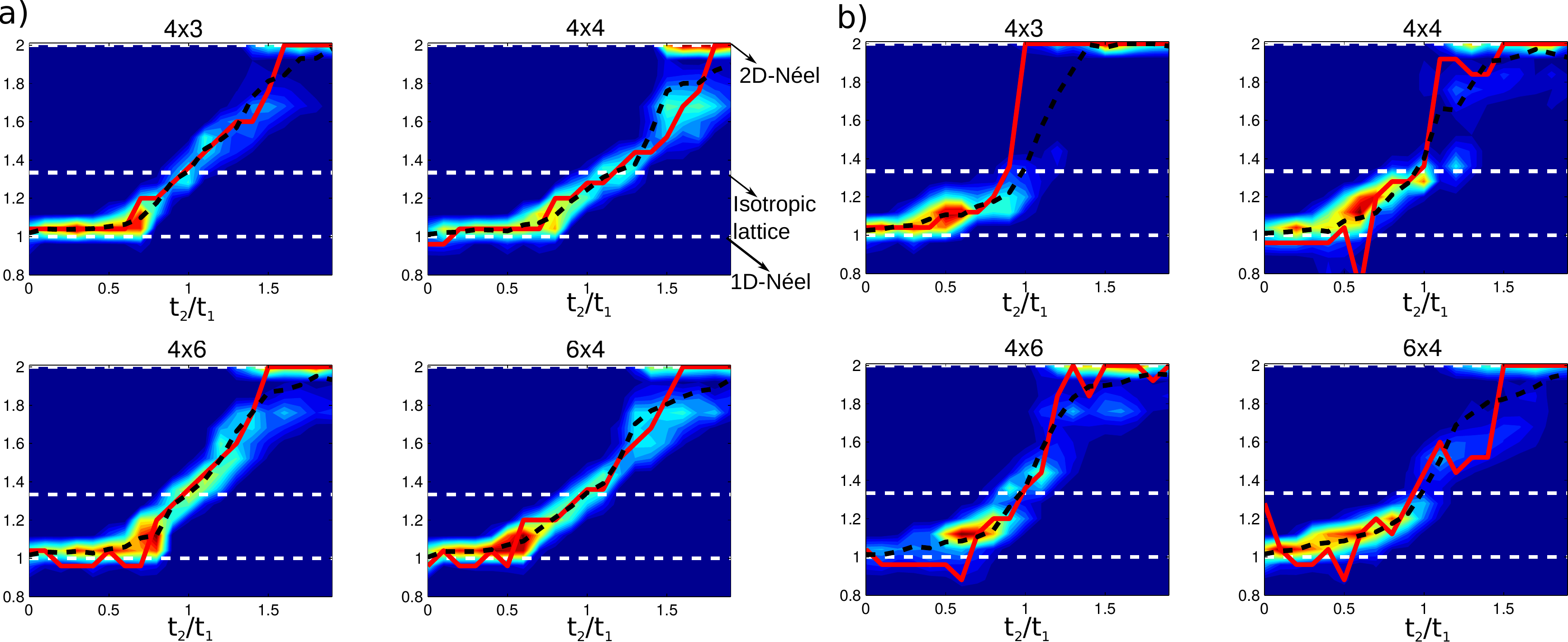}
\caption{Color map distribution of the ordering vector, $Q_x/\pi$, for a sample of 200 set of random phases $\lbrace\varphi_{i,j}\rbrace$ defined as RBC (RRBC) in panels a) (b)), as a function of the anisotropy $t_{2}/t_{1}$. Horizontal lines show the classical values for the ordered phases, $Q_{x}/\pi=1, 4/3 $ and $2$ corresponding to 1D-N\'{e}el, isotropic lattice and 2D-N\'{e}el respectively. The black dashed line depicts $\langle Q_x \rangle_d$, i.e., the averaged value over the disorder, red solid line depicts $Q_x^{min}$ corresponding to the ground state with minimum energy $\vert\Psi_{0,min}\rangle$.}
\label{fig:Qx}
\end{figure*}
The next quantity we look at is $Q_x$ (at $Q_y=0$).
In Fig.\ref{fig:Qx}, we show both the distribution of $Q_{x}$ (color map) as well as its averaged value over the disorder (dashed black line). 
Ordered phases correspond to $Q_x/\pi=1, 4/3$, and $2$  for  the 1D-N\'{e}el, isotropic lattice and 2D-N\'{e}el orders respectively. ED studies with usual boundary conditions show that the $Q_x$ smoothly changes from $t_{2}/t_{1}=0$ up to $t_{2}/t_{1}\simeq 1.6$ where it abruptly jumps to the value $2\pi$  signaling the transition to the 2D intrachain N\'eel order. Such studies as well as MSWT do not show any signature of a SL for  $t_{2}/t_{1} \sim 0.5$.   Our results show that, as expected for $t_{2}>0$, the system is not anymore in the uncoupled chain limit and $Q_x$ increases over the classical value 1. At values $t_{2}/t_{1}\sim 0.5, 1.4$  we observe that distribution of $Q_{x}$ spreads significantly.  This translates, as we will see later, into a broader filling of the first Brillouin zone, since a wide range of k-vectors are in this regions allowed. At the isotropic point, $t_{2}/t_{1}=1$, and only there, $Q_{x}/\pi=4/3$ while the ordering vector smoothly changes along the spiral phase. All the features of $Q_{x}$ are quite independent on the size and geometry of the plaquette, as it can be seen by inspection of the figure, as in the way in which randomness is introduced, RBC panels a) and RRBC panels b). To complete the study, we also display $Q_{x}^{min}$ (thick red line) corresponding to the ordering vector associated to the lowest ground state energy, $E_{0,min}$, as in the study of Ref.\cite{Thesberg2014}. As expected, such quantity has a more pronounced dependence on the lattice geometry and size. 

\begin{figure}
\includegraphics[width=0.8\columnwidth]{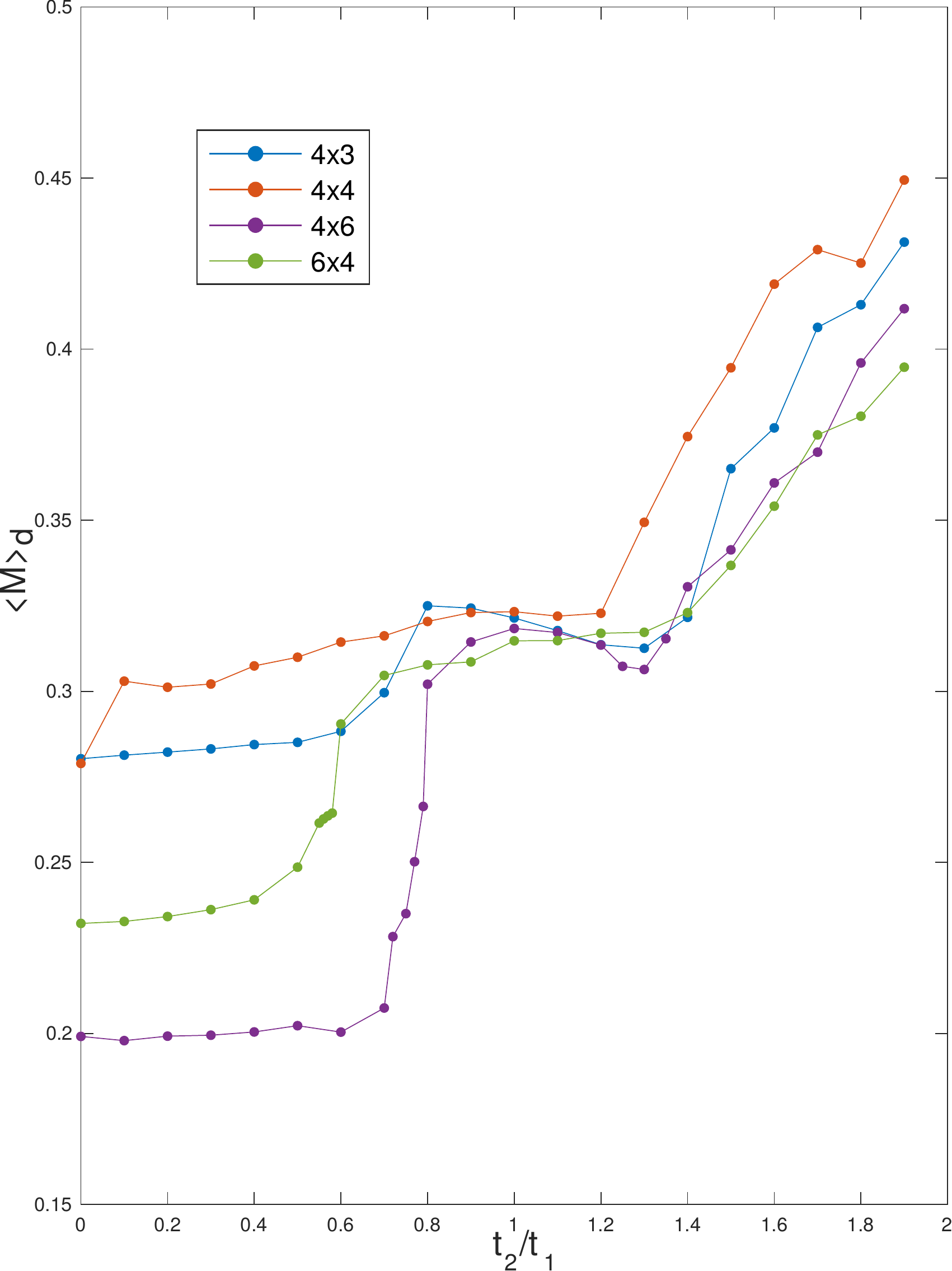}
\caption{Order parameter, $\langle M\rangle_d$, averaged over the disorder as a function of $t_2/t_1$ for different lattice sizes and geometries. The results are obtained using RBC.}
\label{fig:orderparameter}
\end{figure}

The occurrence of a quantum SL phase, in contrast to ordered phases, should be also reflected by the absence of LRO and thus, a comparatively smaller value of the order parameter $M$. In 
Fig. \ref{fig:orderparameter}, we plot $\langle M\rangle_d$ for different geometries and lattice sizes. For all of them a dip appears at  $t_{2}/t_{1}\sim1.3$ while for the 4x6 lattice we can also observe a dip around $t_{2}/t_{1}\sim 0.6$. 

\begin{figure*}[t]
\includegraphics[width=1.5\columnwidth]{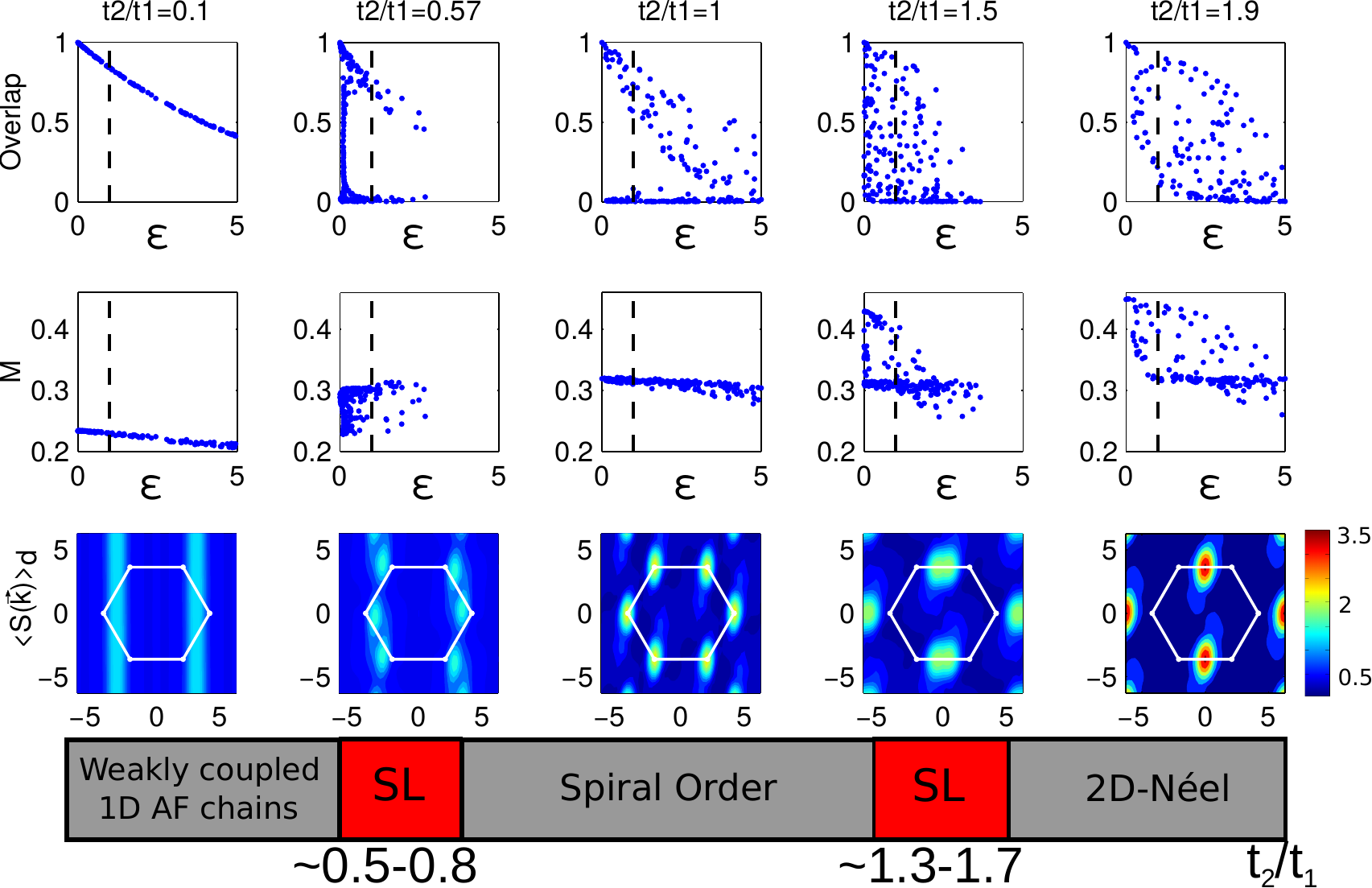}
\caption{First row: overlap between the ground state which minimizes the energy for a fixed $t_{2}/t_{1}$, $|\Psi_{0,min}\rangle$, and the ground states obtained for the other random configurations of the sample, $|\langle \Psi_{0,min}|\Psi_{0,i}(\phi_{1},\phi_{2})\rangle|$, as a function of the renormalized energy, $\epsilon$. The vertical dashed line indicates the bias on the energy set to select from the random sample of boundary conditions only those ground states which are energetically closed to the lowest one, with energies which are not larger than $1\%$ of $E_{0,min}$ ($\epsilon<1$).
Second row: Order parameter, $M$, as defined in the text. In the ordered phases, $M$ remains the same for all energetically close configurations, for quantum disordered systems this is not the case showing that an order parameter can be defined on average but it is a meaningless quantity. 
Third row: static spin structure factor, $\langle S(\vec{k})\rangle_d$,  averaged over the configurations fulfiling $\epsilon<1$. Notice that, for the ordered phases, $\langle S(\vec{k})\rangle_d$ is as expected and for the disordered ones the maxima are blurred.
Fourth row: quantum phase diagram obtained from RBC with two possible SL phases between the ordered ones.}
\label{fig:overlap}
\end{figure*}
The effect of random boundary conditions to study disordered quantum spin phases can be summarized in Fig.\ref{fig:overlap} where we concentrate most of our results taking as a representative case a plaquette of  $6\times 4$ sites. At the top of the figure we display the overlap  between the ground state obtained for each random boundary configuration, $|\Psi_{0,i} (\phi_{1},\phi_{2}) \rangle$,  with the ground state 
$|\Psi_{0,min}\rangle$ corresponding to the configuration with $E_{0,min}$ as a function of $\epsilon$ for some the selected values of $t_{2}/t_{1}$. Note that when plotting as a function of $\epsilon$, we are ordering by increasing energy. The vertical line indicates which ground states we retain from the sampling to perform the averages over the disorder. For $0<t_{2}/t_{1}\leq 0.50$  the overlap smoothly decreases as the energy of the configurations increases. Nevertheless, all ground state configurations close in energy to $|\Psi_{0,min}\rangle$ have the same M and $\langle S(\vec{k})\rangle_d$ corresponds to weakly coupled 1D N\'eel chains. As depicted in the second column, for $t_{2}/t_{1}=0.57$ a drastic change appears. Many different configurations are quasi degenerated in energy, but their corresponding ground states can be very different as indicated by all possible values of the overlap. The order parameter of these quasi-degenerated configurations span all possible values between the M associated to 1D N\'eel chains and the one associated to spiral ordering. 
Interestingly enough, the ground state energy of any random configuration does not deviate more than $\sim3\%$ of the minimal one. These features are compatible with a quantum SL, a disordered system with a large variety of superposed  ground states, as for instance is a resonating valence bond states (RVB). The associated $\langle S(\vec{k})\rangle_d$ is depicted at the bottom. Our calculations show that this phenomena can persist till $t_{2}/t_{1}\sim 0.8$. The precise border depends on the size/geometry of the lattice. At the spiral phase, here depicted in the third column by its most representative case, $t_{2}/t_{1}=1$, two almost degenerated orthogonal ground states with minimal energy appear. They correspond to the two well known chiral ground states known to exist in the spiral phase. Two branches of ground states configurations appear close on energy, the upper one has an overlap $\langle \Psi_{0,min}|\Psi_{0,i}(\phi_{1},\phi_{2})\rangle\simeq 1$ if they share the same chirality and zero if they correspond to different chirality. The order parameter, $M$, attains the same value independently of the chirality, stressing thus the character of the ordered phase. For $t_2/t_1\in [1.3, 1,7]$ similar features as
in $t_2/t_1\sim 0.6$ appear. Again, a large number of configurations are quasi-degenerated on energy. The overlap between all compatible ground states runs between 1 and zero. The order parameter $M$ of the configurations decreases as compared to the spiral phase for $t_2/t_1=1.3$ and afterwards it is not well defined as it spans over a broad range of values. We depict the behavior of such phase at $t_2/t_1=1.5$ where again a large number of configurations quasi degenerated on energy appear. In a similar fashion as it happens for $t_2/t_1\sim 0.6$, there is not a defined value of $M$ and the corresponding $\langle S(\vec{k})\rangle_d$ shows a Brillouin zone with broad maxima if compared to the ordered phases. Finally, for  $t_2/t_1>1.7$ we approximately recover the results of the 2D N\'eel order, showing the tendency of selecting a single ground state with minimal energy, a well defined M and the familiar $S(\vec{k})$ of the rectangular lattice.  We also add at the bottom of the plot an approximate phase diagram obtained from the results of our study, indicating the presence of two regions compatible with a gapped SL. Let us finally comment about the dependence of the results on the used cluster. Our results are general in the sense that the two regions characterized by a large quasi-degenerate energy ground states persist in all cases. However, the precise location at which such degeneracy appears and how long it extends depends on the lattice geometry and size.
\section{Conclusions}
\label{sec:conclusions}
We have approached quantum AF spin systems using ED with random boundary conditions. In this work, we have concentrated our efforts in the spin-1/2 XY model in the SATL, aiming at obtaining a signature of the predicted quantum spin liquids phases. Our results show that there are regions of the phase diagram where many different ground states (sharing the same energy) are compatible. In these regions, the associated observables (i.e. the spin structure factor, ordering vector and order parameter) are, however, very different. The location of these regions, that we identify as spin liquids, agree closely with numerical predictions obtained using e.g., PEPS. 
Interestingly enough, our method not only provides significant signatures of the disordered phases, but it is also robust in reproducing the features of the ordered phases, independently of the sample used to simulate random boundaries (if it is large enough), the size of the plaquette and the energy bias used to select random configurations.  It seems reasonable to conclude that the ordered phases are 
robust in front of quantum fluctuations while the latter are clearly enhanced in the disordered phases. 
 Finally, it is worth to mention that the method we propose is completely general and can be applied for any 2D quantum spin disordered system not only with ED but any numerical method relying on boundary conditions. \\
\textit{Acknowledgements}.  We acknowledge financial support from the Spanish MINECO projects FIS2013-40627-P, the Generalitat de Catalunya CIRIT (2014-SGR-966, 2014-SGR-874).\\
\bibliographystyle{apsrev4-1}
\bibliography{biblio_RBC}

\begin{thebibliography}{37}%
\makeatletter
\providecommand \@ifxundefined [1]{%
 \@ifx{#1\undefined}
}%
\providecommand \@ifnum [1]{%
 \ifnum #1\expandafter \@firstoftwo
 \else \expandafter \@secondoftwo
 \fi
}%
\providecommand \@ifx [1]{%
 \ifx #1\expandafter \@firstoftwo
 \else \expandafter \@secondoftwo
 \fi
}%
\providecommand \natexlab [1]{#1}%
\providecommand \enquote  [1]{``#1''}%
\providecommand \bibnamefont  [1]{#1}%
\providecommand \bibfnamefont [1]{#1}%
\providecommand \citenamefont [1]{#1}%
\providecommand \href@noop [0]{\@secondoftwo}%
\providecommand \href [0]{\begingroup \@sanitize@url \@href}%
\providecommand \@href[1]{\@@startlink{#1}\@@href}%
\providecommand \@@href[1]{\endgroup#1\@@endlink}%
\providecommand \@sanitize@url [0]{\catcode `\\12\catcode `\$12\catcode
  `\&12\catcode `\#12\catcode `\^12\catcode `\_12\catcode `\%12\relax}%
\providecommand \@@startlink[1]{}%
\providecommand \@@endlink[0]{}%
\providecommand \url  [0]{\begingroup\@sanitize@url \@url }%
\providecommand \@url [1]{\endgroup\@href {#1}{\urlprefix }}%
\providecommand \urlprefix  [0]{URL }%
\providecommand \Eprint [0]{\href }%
\providecommand \doibase [0]{http://dx.doi.org/}%
\providecommand \selectlanguage [0]{\@gobble}%
\providecommand \bibinfo  [0]{\@secondoftwo}%
\providecommand \bibfield  [0]{\@secondoftwo}%
\providecommand \translation [1]{[#1]}%
\providecommand \BibitemOpen [0]{}%
\providecommand \bibitemStop [0]{}%
\providecommand \bibitemNoStop [0]{.\EOS\space}%
\providecommand \EOS [0]{\spacefactor3000\relax}%
\providecommand \BibitemShut  [1]{\csname bibitem#1\endcsname}%
\let\auto@bib@innerbib\@empty
\bibitem [{\citenamefont {Misguich}\ and\ \citenamefont
  {Lhuillier}(2012)}]{Lhuillier2012}%
  \BibitemOpen
  \bibfield  {author} {\bibinfo {author} {\bibfnamefont {G.}~\bibnamefont
  {Misguich}}\ and\ \bibinfo {author} {\bibfnamefont {C.}~\bibnamefont
  {Lhuillier}},\ }\enquote {\bibinfo {title} {Two-dimensional quantum
  antiferromagnets},}\ in\ \href {\doibase 10.1142/9789812567819_0005} {\emph
  {\bibinfo {booktitle} {Frustrated Spin Systems}}}\ (\bibinfo  {publisher}
  {World Scientific},\ \bibinfo {year} {2012})\ pp.\ \bibinfo {pages}
  {229--306}\BibitemShut {NoStop}%
\bibitem [{\citenamefont {Kitaev}\ and\ \citenamefont
  {Preskill}(2006)}]{Kitaev2006}%
  \BibitemOpen
  \bibfield  {author} {\bibinfo {author} {\bibfnamefont {A.}~\bibnamefont
  {Kitaev}}\ and\ \bibinfo {author} {\bibfnamefont {J.}~\bibnamefont
  {Preskill}},\ }\href {\doibase 10.1103/PhysRevLett.96.110404} {\bibfield
  {journal} {\bibinfo  {journal} {Phys. Rev. Lett.}\ }\textbf {\bibinfo
  {volume} {96}},\ \bibinfo {pages} {110404} (\bibinfo {year}
  {2006})}\BibitemShut {NoStop}%
\bibitem [{\citenamefont {Levin}\ and\ \citenamefont {Wen}(2006)}]{Levin2006}%
  \BibitemOpen
  \bibfield  {author} {\bibinfo {author} {\bibfnamefont {M.}~\bibnamefont
  {Levin}}\ and\ \bibinfo {author} {\bibfnamefont {X.-G.}\ \bibnamefont
  {Wen}},\ }\href {\doibase 10.1103/PhysRevLett.96.110405} {\bibfield
  {journal} {\bibinfo  {journal} {Phys. Rev. Lett.}\ }\textbf {\bibinfo
  {volume} {96}},\ \bibinfo {pages} {110405} (\bibinfo {year}
  {2006})}\BibitemShut {NoStop}%
\bibitem [{\citenamefont {Isakov}\ \emph {et~al.}(2011)\citenamefont {Isakov},
  \citenamefont {Hastings},\ and\ \citenamefont {Melko}}]{Isakov2011}%
  \BibitemOpen
  \bibfield  {author} {\bibinfo {author} {\bibfnamefont {S.~V.}\ \bibnamefont
  {Isakov}}, \bibinfo {author} {\bibfnamefont {M.~B.}\ \bibnamefont
  {Hastings}}, \ and\ \bibinfo {author} {\bibfnamefont {R.~G.}\ \bibnamefont
  {Melko}},\ }\href {\doibase 10.1038/nphys2036} {\bibfield  {journal}
  {\bibinfo  {journal} {Nature Phys.}\ }\textbf {\bibinfo {volume} {7}},\
  \bibinfo {pages} {772} (\bibinfo {year} {2011})}\BibitemShut {NoStop}%
\bibitem [{\citenamefont {Wen}(2016)}]{Wen2016}%
  \BibitemOpen
  \bibfield  {author} {\bibinfo {author} {\bibfnamefont {X.-G.}\ \bibnamefont
  {Wen}},\ }\href {http://arxiv.org/abs/1610.03911} {\  (\bibinfo {year}
  {2016})},\ \Eprint {http://arxiv.org/abs/1610.03911} {arXiv:1610.03911}
  \BibitemShut {NoStop}%
\bibitem [{\citenamefont {Balents}(2010)}]{Balents2010}%
  \BibitemOpen
  \bibfield  {author} {\bibinfo {author} {\bibfnamefont {L.}~\bibnamefont
  {Balents}},\ }\href {http://dx.doi.org/10.1038/nature08917} {\bibfield
  {journal} {\bibinfo  {journal} {Nature}\ }\textbf {\bibinfo {volume} {464}},\
  \bibinfo {pages} {199} (\bibinfo {year} {2010})}\BibitemShut {NoStop}%
\bibitem [{\citenamefont {Kitaev}(2006)}]{Kitaev2006a}%
  \BibitemOpen
  \bibfield  {author} {\bibinfo {author} {\bibfnamefont {A.}~\bibnamefont
  {Kitaev}},\ }\href {http://dx.doi.org/10.1016/j.aop.2005.10.005} {\bibfield
  {journal} {\bibinfo  {journal} {Ann. Phys. (Leipz.)}\ }\textbf {\bibinfo
  {volume} {321}},\ \bibinfo {pages} {2} (\bibinfo {year} {2006})}\BibitemShut
  {NoStop}%
\bibitem [{\citenamefont {Moessner}\ and\ \citenamefont
  {Sondhi}(2001)}]{Moessner2001}%
  \BibitemOpen
  \bibfield  {author} {\bibinfo {author} {\bibfnamefont {R.}~\bibnamefont
  {Moessner}}\ and\ \bibinfo {author} {\bibfnamefont {S.~L.}\ \bibnamefont
  {Sondhi}},\ }\href {\doibase 10.1103/PhysRevLett.86.1881} {\bibfield
  {journal} {\bibinfo  {journal} {Phys. Rev. Lett.}\ }\textbf {\bibinfo
  {volume} {86}},\ \bibinfo {pages} {1881} (\bibinfo {year}
  {2001})}\BibitemShut {NoStop}%
\bibitem [{\citenamefont {Yamashita}\ \emph {et~al.}(2010)\citenamefont
  {Yamashita}, \citenamefont {Nakata}, \citenamefont {Senshu}, \citenamefont
  {Nagata}, \citenamefont {Yamamoto}, \citenamefont {Kato}, \citenamefont
  {Shibauchi},\ and\ \citenamefont {Matsuda}}]{Yamashita2010}%
  \BibitemOpen
  \bibfield  {author} {\bibinfo {author} {\bibfnamefont {M.}~\bibnamefont
  {Yamashita}}, \bibinfo {author} {\bibfnamefont {N.}~\bibnamefont {Nakata}},
  \bibinfo {author} {\bibfnamefont {Y.}~\bibnamefont {Senshu}}, \bibinfo
  {author} {\bibfnamefont {M.}~\bibnamefont {Nagata}}, \bibinfo {author}
  {\bibfnamefont {H.~M.}\ \bibnamefont {Yamamoto}}, \bibinfo {author}
  {\bibfnamefont {R.}~\bibnamefont {Kato}}, \bibinfo {author} {\bibfnamefont
  {T.}~\bibnamefont {Shibauchi}}, \ and\ \bibinfo {author} {\bibfnamefont
  {Y.}~\bibnamefont {Matsuda}},\ }\href {\doibase 10.1126/science.1188200}
  {\bibfield  {journal} {\bibinfo  {journal} {Science}\ }\textbf {\bibinfo
  {volume} {328}},\ \bibinfo {pages} {1246} (\bibinfo {year}
  {2010})}\BibitemShut {NoStop}%
\bibitem [{\citenamefont {Yamashita}(2008)}]{Yamashita2008}%
  \BibitemOpen
  \bibfield  {author} {\bibinfo {author} {\bibfnamefont {S.}~\bibnamefont
  {Yamashita}},\ }\href {http://dx.doi.org/10.1038/nphys942} {\bibfield
  {journal} {\bibinfo  {journal} {Nature Phys.}\ }\textbf {\bibinfo {volume}
  {4}},\ \bibinfo {pages} {459} (\bibinfo {year} {2008})}\BibitemShut {NoStop}%
\bibitem [{\citenamefont {de~Vries}\ \emph {et~al.}(2009)\citenamefont
  {de~Vries}, \citenamefont {Stewart}, \citenamefont {Deen}, \citenamefont
  {Piatek}, \citenamefont {Nilsen}, \citenamefont {R\o{}nnow},\ and\
  \citenamefont {Harrison}}]{DeVries2009}%
  \BibitemOpen
  \bibfield  {author} {\bibinfo {author} {\bibfnamefont {M.~A.}\ \bibnamefont
  {de~Vries}}, \bibinfo {author} {\bibfnamefont {J.~R.}\ \bibnamefont
  {Stewart}}, \bibinfo {author} {\bibfnamefont {P.~P.}\ \bibnamefont {Deen}},
  \bibinfo {author} {\bibfnamefont {J.~O.}\ \bibnamefont {Piatek}}, \bibinfo
  {author} {\bibfnamefont {G.~J.}\ \bibnamefont {Nilsen}}, \bibinfo {author}
  {\bibfnamefont {H.~M.}\ \bibnamefont {R\o{}nnow}}, \ and\ \bibinfo {author}
  {\bibfnamefont {A.}~\bibnamefont {Harrison}},\ }\href {\doibase
  10.1103/PhysRevLett.103.237201} {\bibfield  {journal} {\bibinfo  {journal}
  {Phys. Rev. Lett.}\ }\textbf {\bibinfo {volume} {103}},\ \bibinfo {pages}
  {237201} (\bibinfo {year} {2009})}\BibitemShut {NoStop}%
\bibitem [{\citenamefont {Shimizu}\ \emph {et~al.}(2003)\citenamefont
  {Shimizu}, \citenamefont {Miyagawa}, \citenamefont {Kanoda}, \citenamefont
  {Maesato},\ and\ \citenamefont {Saito}}]{Shimizu2003}%
  \BibitemOpen
  \bibfield  {author} {\bibinfo {author} {\bibfnamefont {Y.}~\bibnamefont
  {Shimizu}}, \bibinfo {author} {\bibfnamefont {K.}~\bibnamefont {Miyagawa}},
  \bibinfo {author} {\bibfnamefont {K.}~\bibnamefont {Kanoda}}, \bibinfo
  {author} {\bibfnamefont {M.}~\bibnamefont {Maesato}}, \ and\ \bibinfo
  {author} {\bibfnamefont {G.}~\bibnamefont {Saito}},\ }\href {\doibase
  10.1103/PhysRevLett.91.107001} {\bibfield  {journal} {\bibinfo  {journal}
  {Phys. Rev. Lett.}\ }\textbf {\bibinfo {volume} {91}},\ \bibinfo {pages}
  {107001} (\bibinfo {year} {2003})}\BibitemShut {NoStop}%
\bibitem [{\citenamefont {Mendels}\ \emph {et~al.}(2007)\citenamefont
  {Mendels}, \citenamefont {Bert}, \citenamefont {de~Vries}, \citenamefont
  {Olariu}, \citenamefont {Harrison}, \citenamefont {Duc}, \citenamefont
  {Trombe}, \citenamefont {Lord}, \citenamefont {Amato},\ and\ \citenamefont
  {Baines}}]{Mendels2007}%
  \BibitemOpen
  \bibfield  {author} {\bibinfo {author} {\bibfnamefont {P.}~\bibnamefont
  {Mendels}}, \bibinfo {author} {\bibfnamefont {F.}~\bibnamefont {Bert}},
  \bibinfo {author} {\bibfnamefont {M.~A.}\ \bibnamefont {de~Vries}}, \bibinfo
  {author} {\bibfnamefont {A.}~\bibnamefont {Olariu}}, \bibinfo {author}
  {\bibfnamefont {A.}~\bibnamefont {Harrison}}, \bibinfo {author}
  {\bibfnamefont {F.}~\bibnamefont {Duc}}, \bibinfo {author} {\bibfnamefont
  {J.~C.}\ \bibnamefont {Trombe}}, \bibinfo {author} {\bibfnamefont {J.~S.}\
  \bibnamefont {Lord}}, \bibinfo {author} {\bibfnamefont {A.}~\bibnamefont
  {Amato}}, \ and\ \bibinfo {author} {\bibfnamefont {C.}~\bibnamefont
  {Baines}},\ }\href {\doibase 10.1103/PhysRevLett.98.077204} {\bibfield
  {journal} {\bibinfo  {journal} {Phys. Rev. Lett.}\ }\textbf {\bibinfo
  {volume} {98}},\ \bibinfo {pages} {077204} (\bibinfo {year}
  {2007})}\BibitemShut {NoStop}%
\bibitem [{\citenamefont {Kurosaki}\ \emph {et~al.}(2005)\citenamefont
  {Kurosaki}, \citenamefont {Shimizu}, \citenamefont {Miyagawa}, \citenamefont
  {Kanoda},\ and\ \citenamefont {Saito}}]{Kurosaki2005}%
  \BibitemOpen
  \bibfield  {author} {\bibinfo {author} {\bibfnamefont {Y.}~\bibnamefont
  {Kurosaki}}, \bibinfo {author} {\bibfnamefont {Y.}~\bibnamefont {Shimizu}},
  \bibinfo {author} {\bibfnamefont {K.}~\bibnamefont {Miyagawa}}, \bibinfo
  {author} {\bibfnamefont {K.}~\bibnamefont {Kanoda}}, \ and\ \bibinfo {author}
  {\bibfnamefont {G.}~\bibnamefont {Saito}},\ }\href {\doibase
  10.1103/PhysRevLett.95.177001} {\bibfield  {journal} {\bibinfo  {journal}
  {Phys. Rev. Lett.}\ }\textbf {\bibinfo {volume} {95}},\ \bibinfo {pages}
  {177001} (\bibinfo {year} {2005})}\BibitemShut {NoStop}%
\bibitem [{\citenamefont {Itou}\ \emph {et~al.}(2008)\citenamefont {Itou},
  \citenamefont {Oyamada}, \citenamefont {Maegawa}, \citenamefont {Tamura},\
  and\ \citenamefont {Kato}}]{Itou2008}%
  \BibitemOpen
  \bibfield  {author} {\bibinfo {author} {\bibfnamefont {T.}~\bibnamefont
  {Itou}}, \bibinfo {author} {\bibfnamefont {A.}~\bibnamefont {Oyamada}},
  \bibinfo {author} {\bibfnamefont {S.}~\bibnamefont {Maegawa}}, \bibinfo
  {author} {\bibfnamefont {M.}~\bibnamefont {Tamura}}, \ and\ \bibinfo {author}
  {\bibfnamefont {R.}~\bibnamefont {Kato}},\ }\href {\doibase
  10.1103/PhysRevB.77.104413} {\bibfield  {journal} {\bibinfo  {journal} {Phys.
  Rev. B}\ }\textbf {\bibinfo {volume} {77}},\ \bibinfo {pages} {104413}
  (\bibinfo {year} {2008})}\BibitemShut {NoStop}%
\bibitem [{\citenamefont {Helton}\ \emph {et~al.}(2007)\citenamefont {Helton},
  \citenamefont {Matan}, \citenamefont {Shores}, \citenamefont {Nytko},
  \citenamefont {Bartlett}, \citenamefont {Yoshida}, \citenamefont {Takano},
  \citenamefont {Suslov}, \citenamefont {Qiu}, \citenamefont {Chung},
  \citenamefont {Nocera},\ and\ \citenamefont {Lee}}]{Helton2007}%
  \BibitemOpen
  \bibfield  {author} {\bibinfo {author} {\bibfnamefont {J.~S.}\ \bibnamefont
  {Helton}}, \bibinfo {author} {\bibfnamefont {K.}~\bibnamefont {Matan}},
  \bibinfo {author} {\bibfnamefont {M.~P.}\ \bibnamefont {Shores}}, \bibinfo
  {author} {\bibfnamefont {E.~A.}\ \bibnamefont {Nytko}}, \bibinfo {author}
  {\bibfnamefont {B.~M.}\ \bibnamefont {Bartlett}}, \bibinfo {author}
  {\bibfnamefont {Y.}~\bibnamefont {Yoshida}}, \bibinfo {author} {\bibfnamefont
  {Y.}~\bibnamefont {Takano}}, \bibinfo {author} {\bibfnamefont
  {A.}~\bibnamefont {Suslov}}, \bibinfo {author} {\bibfnamefont
  {Y.}~\bibnamefont {Qiu}}, \bibinfo {author} {\bibfnamefont {J.-H.}\
  \bibnamefont {Chung}}, \bibinfo {author} {\bibfnamefont {D.~G.}\ \bibnamefont
  {Nocera}}, \ and\ \bibinfo {author} {\bibfnamefont {Y.~S.}\ \bibnamefont
  {Lee}},\ }\href {\doibase 10.1103/PhysRevLett.98.107204} {\bibfield
  {journal} {\bibinfo  {journal} {Phys. Rev. Lett.}\ }\textbf {\bibinfo
  {volume} {98}},\ \bibinfo {pages} {107204} (\bibinfo {year}
  {2007})}\BibitemShut {NoStop}%
\bibitem [{\citenamefont {Han}\ \emph {et~al.}(2012)\citenamefont {Han},
  \citenamefont {Helton}, \citenamefont {Chu}, \citenamefont {Nocera},
  \citenamefont {Rodriguez-Rivera}, \citenamefont {Broholm},\ and\
  \citenamefont {Lee}}]{Han2012}%
  \BibitemOpen
  \bibfield  {author} {\bibinfo {author} {\bibfnamefont {T.-H.}\ \bibnamefont
  {Han}}, \bibinfo {author} {\bibfnamefont {J.~S.}\ \bibnamefont {Helton}},
  \bibinfo {author} {\bibfnamefont {S.}~\bibnamefont {Chu}}, \bibinfo {author}
  {\bibfnamefont {D.~G.}\ \bibnamefont {Nocera}}, \bibinfo {author}
  {\bibfnamefont {J.~a.}\ \bibnamefont {Rodriguez-Rivera}}, \bibinfo {author}
  {\bibfnamefont {C.}~\bibnamefont {Broholm}}, \ and\ \bibinfo {author}
  {\bibfnamefont {Y.~S.}\ \bibnamefont {Lee}},\ }\href {\doibase
  10.1038/nature11659} {\bibfield  {journal} {\bibinfo  {journal} {Nature}\
  }\textbf {\bibinfo {volume} {492}},\ \bibinfo {pages} {406} (\bibinfo {year}
  {2012})}\BibitemShut {NoStop}%
\bibitem [{\citenamefont {F\aa{}k}\ \emph {et~al.}(2012)\citenamefont
  {F\aa{}k}, \citenamefont {Kermarrec}, \citenamefont {Messio}, \citenamefont
  {Bernu}, \citenamefont {Lhuillier}, \citenamefont {Bert}, \citenamefont
  {Mendels}, \citenamefont {Koteswararao}, \citenamefont {Bouquet},
  \citenamefont {Ollivier}, \citenamefont {Hillier}, \citenamefont {Amato},
  \citenamefont {Colman},\ and\ \citenamefont {Wills}}]{Fak2012}%
  \BibitemOpen
  \bibfield  {author} {\bibinfo {author} {\bibfnamefont {B.}~\bibnamefont
  {F\aa{}k}}, \bibinfo {author} {\bibfnamefont {E.}~\bibnamefont {Kermarrec}},
  \bibinfo {author} {\bibfnamefont {L.}~\bibnamefont {Messio}}, \bibinfo
  {author} {\bibfnamefont {B.}~\bibnamefont {Bernu}}, \bibinfo {author}
  {\bibfnamefont {C.}~\bibnamefont {Lhuillier}}, \bibinfo {author}
  {\bibfnamefont {F.}~\bibnamefont {Bert}}, \bibinfo {author} {\bibfnamefont
  {P.}~\bibnamefont {Mendels}}, \bibinfo {author} {\bibfnamefont
  {B.}~\bibnamefont {Koteswararao}}, \bibinfo {author} {\bibfnamefont
  {F.}~\bibnamefont {Bouquet}}, \bibinfo {author} {\bibfnamefont
  {J.}~\bibnamefont {Ollivier}}, \bibinfo {author} {\bibfnamefont {A.~D.}\
  \bibnamefont {Hillier}}, \bibinfo {author} {\bibfnamefont {A.}~\bibnamefont
  {Amato}}, \bibinfo {author} {\bibfnamefont {R.~H.}\ \bibnamefont {Colman}}, \
  and\ \bibinfo {author} {\bibfnamefont {A.~S.}\ \bibnamefont {Wills}},\ }\href
  {\doibase 10.1103/PhysRevLett.109.037208} {\bibfield  {journal} {\bibinfo
  {journal} {Phys. Rev. Lett.}\ }\textbf {\bibinfo {volume} {109}},\ \bibinfo
  {pages} {037208} (\bibinfo {year} {2012})}\BibitemShut {NoStop}%
\bibitem [{\citenamefont {Coldea}\ \emph {et~al.}(2001)\citenamefont {Coldea},
  \citenamefont {Tennant}, \citenamefont {Tsvelik},\ and\ \citenamefont
  {Tylczynski}}]{Coldea2001}%
  \BibitemOpen
  \bibfield  {author} {\bibinfo {author} {\bibfnamefont {R.}~\bibnamefont
  {Coldea}}, \bibinfo {author} {\bibfnamefont {D.~A.}\ \bibnamefont {Tennant}},
  \bibinfo {author} {\bibfnamefont {A.~M.}\ \bibnamefont {Tsvelik}}, \ and\
  \bibinfo {author} {\bibfnamefont {Z.}~\bibnamefont {Tylczynski}},\ }\href
  {\doibase 10.1103/PhysRevLett.86.1335} {\bibfield  {journal} {\bibinfo
  {journal} {Phys. Rev. Lett.}\ }\textbf {\bibinfo {volume} {86}},\ \bibinfo
  {pages} {1335} (\bibinfo {year} {2001})}\BibitemShut {NoStop}%
\bibitem [{\citenamefont {Struck}\ \emph {et~al.}(2011)\citenamefont {Struck},
  \citenamefont {{\"{O}}lschl{\"{a}}ger}, \citenamefont {{Le Targat}},
  \citenamefont {Soltan-Panahi}, \citenamefont {Eckardt}, \citenamefont
  {Lewenstein}, \citenamefont {Windpassinger},\ and\ \citenamefont
  {Sengstock}}]{Struck2011}%
  \BibitemOpen
  \bibfield  {author} {\bibinfo {author} {\bibfnamefont {J.}~\bibnamefont
  {Struck}}, \bibinfo {author} {\bibfnamefont {C.}~\bibnamefont
  {{\"{O}}lschl{\"{a}}ger}}, \bibinfo {author} {\bibfnamefont {R.}~\bibnamefont
  {{Le Targat}}}, \bibinfo {author} {\bibfnamefont {P.}~\bibnamefont
  {Soltan-Panahi}}, \bibinfo {author} {\bibfnamefont {A.}~\bibnamefont
  {Eckardt}}, \bibinfo {author} {\bibfnamefont {M.}~\bibnamefont {Lewenstein}},
  \bibinfo {author} {\bibfnamefont {P.}~\bibnamefont {Windpassinger}}, \ and\
  \bibinfo {author} {\bibfnamefont {K.}~\bibnamefont {Sengstock}},\ }\href
  {http://science.sciencemag.org/content/333/6045/996.abstract} {\bibfield
  {journal} {\bibinfo  {journal} {Science}\ }\textbf {\bibinfo {volume}
  {333}},\ \bibinfo {pages} {996 LP } (\bibinfo {year} {2011})}\BibitemShut
  {NoStop}%
\bibitem [{\citenamefont {Hauke}(2013)}]{Hauke2013a}%
  \BibitemOpen
  \bibfield  {author} {\bibinfo {author} {\bibfnamefont {P.}~\bibnamefont
  {Hauke}},\ }\href {\doibase 10.1103/PhysRevB.87.014415} {\bibfield  {journal}
  {\bibinfo  {journal} {Phys. Rev. B}\ }\textbf {\bibinfo {volume} {87}},\
  \bibinfo {pages} {014415} (\bibinfo {year} {2013})}\BibitemShut {NoStop}%
\bibitem [{\citenamefont {Schmied}\ \emph {et~al.}(2010)\citenamefont
  {Schmied}, \citenamefont {Roscilde}, \citenamefont {Murg}, \citenamefont
  {Cirac},\ and\ \citenamefont {Roman}}]{Hauke2010}%
  \BibitemOpen
  \bibfield  {author} {\bibinfo {author} {\bibfnamefont {P.~H.}\ \bibnamefont
  {Schmied}}, \bibinfo {author} {\bibfnamefont {T.}~\bibnamefont {Roscilde}},
  \bibinfo {author} {\bibfnamefont {V.}~\bibnamefont {Murg}}, \bibinfo {author}
  {\bibfnamefont {J.~I.}\ \bibnamefont {Cirac}}, \ and\ \bibinfo {author}
  {\bibnamefont {Roman}},\ }\href
  {http://stacks.iop.org/1367-2630/12/i=5/a=053036} {\bibfield  {journal}
  {\bibinfo  {journal} {New J. Phys.}\ }\textbf {\bibinfo {volume} {12}},\
  \bibinfo {pages} {53036} (\bibinfo {year} {2010})}\BibitemShut {NoStop}%
\bibitem [{\citenamefont {Celi}\ \emph {et~al.}(2016)\citenamefont {Celi},
  \citenamefont {Grass}, \citenamefont {Ferris}, \citenamefont {Padhi},
  \citenamefont {Ravent\'os}, \citenamefont {Simonet}, \citenamefont
  {Sengstock},\ and\ \citenamefont {Lewenstein}}]{Celi2016}%
  \BibitemOpen
  \bibfield  {author} {\bibinfo {author} {\bibfnamefont {A.}~\bibnamefont
  {Celi}}, \bibinfo {author} {\bibfnamefont {T.}~\bibnamefont {Grass}},
  \bibinfo {author} {\bibfnamefont {A.~J.}\ \bibnamefont {Ferris}}, \bibinfo
  {author} {\bibfnamefont {B.}~\bibnamefont {Padhi}}, \bibinfo {author}
  {\bibfnamefont {D.}~\bibnamefont {Ravent\'os}}, \bibinfo {author}
  {\bibfnamefont {J.}~\bibnamefont {Simonet}}, \bibinfo {author} {\bibfnamefont
  {K.}~\bibnamefont {Sengstock}}, \ and\ \bibinfo {author} {\bibfnamefont
  {M.}~\bibnamefont {Lewenstein}},\ }\href {\doibase
  10.1103/PhysRevB.94.075110} {\bibfield  {journal} {\bibinfo  {journal} {Phys.
  Rev. B}\ }\textbf {\bibinfo {volume} {94}},\ \bibinfo {pages} {075110}
  (\bibinfo {year} {2016})}\BibitemShut {NoStop}%
\bibitem [{\citenamefont {Yamamoto}\ \emph {et~al.}(2014)\citenamefont
  {Yamamoto}, \citenamefont {Marmorini},\ and\ \citenamefont
  {Danshita}}]{Yamamoto2014}%
  \BibitemOpen
  \bibfield  {author} {\bibinfo {author} {\bibfnamefont {D.}~\bibnamefont
  {Yamamoto}}, \bibinfo {author} {\bibfnamefont {G.}~\bibnamefont {Marmorini}},
  \ and\ \bibinfo {author} {\bibfnamefont {I.}~\bibnamefont {Danshita}},\
  }\href {\doibase 10.1103/PhysRevLett.112.127203} {\bibfield  {journal}
  {\bibinfo  {journal} {Phys. Rev. Lett.}\ }\textbf {\bibinfo {volume} {112}},\
  \bibinfo {pages} {127203} (\bibinfo {year} {2014})}\BibitemShut {NoStop}%
\bibitem [{\citenamefont {Moreno-Cardoner}\ \emph
  {et~al.}(2014{\natexlab{a}})\citenamefont {Moreno-Cardoner}, \citenamefont
  {Perrin}, \citenamefont {Paganelli}, \citenamefont {De~Chiara},\ and\
  \citenamefont {Sanpera}}]{Moreno-Cardoner2014}%
  \BibitemOpen
  \bibfield  {author} {\bibinfo {author} {\bibfnamefont {M.}~\bibnamefont
  {Moreno-Cardoner}}, \bibinfo {author} {\bibfnamefont {H.}~\bibnamefont
  {Perrin}}, \bibinfo {author} {\bibfnamefont {S.}~\bibnamefont {Paganelli}},
  \bibinfo {author} {\bibfnamefont {G.}~\bibnamefont {De~Chiara}}, \ and\
  \bibinfo {author} {\bibfnamefont {A.}~\bibnamefont {Sanpera}},\ }\href
  {\doibase 10.1103/PhysRevB.90.144409} {\bibfield  {journal} {\bibinfo
  {journal} {Phys. Rev. B}\ }\textbf {\bibinfo {volume} {90}},\ \bibinfo
  {pages} {144409} (\bibinfo {year} {2014}{\natexlab{a}})}\BibitemShut
  {NoStop}%
\bibitem [{\citenamefont {Moreno-Cardoner}\ \emph
  {et~al.}(2014{\natexlab{b}})\citenamefont {Moreno-Cardoner}, \citenamefont
  {Paganelli}, \citenamefont {Chiara},\ and\ \citenamefont
  {Sanpera}}]{Moreno-Cardoner2014a}%
  \BibitemOpen
  \bibfield  {author} {\bibinfo {author} {\bibfnamefont {M.}~\bibnamefont
  {Moreno-Cardoner}}, \bibinfo {author} {\bibfnamefont {S.}~\bibnamefont
  {Paganelli}}, \bibinfo {author} {\bibfnamefont {G.~D.}\ \bibnamefont
  {Chiara}}, \ and\ \bibinfo {author} {\bibfnamefont {A.}~\bibnamefont
  {Sanpera}},\ }\href {http://stacks.iop.org/1742-5468/2014/i=10/a=P10008}
  {\bibfield  {journal} {\bibinfo  {journal} {Journal of Statistical Mechanics:
  Theory and Experiment}\ }\textbf {\bibinfo {volume} {2014}},\ \bibinfo
  {pages} {P10008} (\bibinfo {year} {2014}{\natexlab{b}})}\BibitemShut
  {NoStop}%
\bibitem [{\citenamefont {Weng}\ \emph {et~al.}(2006)\citenamefont {Weng},
  \citenamefont {Sheng}, \citenamefont {Weng},\ and\ \citenamefont
  {Bursill}}]{Weng2006}%
  \BibitemOpen
  \bibfield  {author} {\bibinfo {author} {\bibfnamefont {M.~Q.}\ \bibnamefont
  {Weng}}, \bibinfo {author} {\bibfnamefont {D.~N.}\ \bibnamefont {Sheng}},
  \bibinfo {author} {\bibfnamefont {Z.~Y.}\ \bibnamefont {Weng}}, \ and\
  \bibinfo {author} {\bibfnamefont {R.~J.}\ \bibnamefont {Bursill}},\ }\href
  {\doibase 10.1103/PhysRevB.74.012407} {\bibfield  {journal} {\bibinfo
  {journal} {Phys. Rev. B}\ }\textbf {\bibinfo {volume} {74}},\ \bibinfo
  {pages} {772} (\bibinfo {year} {2006})}\BibitemShut {NoStop}%
\bibitem [{\citenamefont {Weichselbaum}\ and\ \citenamefont
  {White}(2011)}]{Weichselbaum2011}%
  \BibitemOpen
  \bibfield  {author} {\bibinfo {author} {\bibfnamefont {A.}~\bibnamefont
  {Weichselbaum}}\ and\ \bibinfo {author} {\bibfnamefont {S.~R.}\ \bibnamefont
  {White}},\ }\href {\doibase 10.1103/PhysRevB.84.245130} {\bibfield  {journal}
  {\bibinfo  {journal} {Phys. Rev. B}\ }\textbf {\bibinfo {volume} {84}},\
  \bibinfo {pages} {245130} (\bibinfo {year} {2011})}\BibitemShut {NoStop}%
\bibitem [{\citenamefont {Hu}\ \emph {et~al.}(2015)\citenamefont {Hu},
  \citenamefont {Gong}, \citenamefont {Zhu},\ and\ \citenamefont
  {Sheng}}]{Hu2015}%
  \BibitemOpen
  \bibfield  {author} {\bibinfo {author} {\bibfnamefont {W.-J.}\ \bibnamefont
  {Hu}}, \bibinfo {author} {\bibfnamefont {S.-S.}\ \bibnamefont {Gong}},
  \bibinfo {author} {\bibfnamefont {W.}~\bibnamefont {Zhu}}, \ and\ \bibinfo
  {author} {\bibfnamefont {D.~N.}\ \bibnamefont {Sheng}},\ }\href {\doibase
  10.1103/PhysRevB.92.140403} {\bibfield  {journal} {\bibinfo  {journal} {Phys.
  Rev. B}\ }\textbf {\bibinfo {volume} {92}},\ \bibinfo {pages} {140403}
  (\bibinfo {year} {2015})}\BibitemShut {NoStop}%
\bibitem [{\citenamefont {Schmied}\ \emph {et~al.}(2008)\citenamefont
  {Schmied}, \citenamefont {Roscilde}, \citenamefont {Murg}, \citenamefont
  {Porras},\ and\ \citenamefont {Cirac}}]{Schmied2008}%
  \BibitemOpen
  \bibfield  {author} {\bibinfo {author} {\bibfnamefont {R.}~\bibnamefont
  {Schmied}}, \bibinfo {author} {\bibfnamefont {T.}~\bibnamefont {Roscilde}},
  \bibinfo {author} {\bibfnamefont {V.}~\bibnamefont {Murg}}, \bibinfo {author}
  {\bibfnamefont {D.}~\bibnamefont {Porras}}, \ and\ \bibinfo {author}
  {\bibfnamefont {J.~I.}\ \bibnamefont {Cirac}},\ }\href
  {http://stacks.iop.org/1367-2630/10/i=4/a=045017} {\bibfield  {journal}
  {\bibinfo  {journal} {New Journal of Physics}\ }\textbf {\bibinfo {volume}
  {10}},\ \bibinfo {pages} {045017} (\bibinfo {year} {2008})}\BibitemShut
  {NoStop}%
\bibitem [{\citenamefont {Tocchio}\ \emph {et~al.}(2009)\citenamefont
  {Tocchio}, \citenamefont {Parola}, \citenamefont {Gros},\ and\ \citenamefont
  {Becca}}]{Tocchio2009}%
  \BibitemOpen
  \bibfield  {author} {\bibinfo {author} {\bibfnamefont {L.~F.}\ \bibnamefont
  {Tocchio}}, \bibinfo {author} {\bibfnamefont {A.}~\bibnamefont {Parola}},
  \bibinfo {author} {\bibfnamefont {C.}~\bibnamefont {Gros}}, \ and\ \bibinfo
  {author} {\bibfnamefont {F.}~\bibnamefont {Becca}},\ }\href {\doibase
  10.1103/PhysRevB.80.064419} {\bibfield  {journal} {\bibinfo  {journal} {Phys.
  Rev. B}\ }\textbf {\bibinfo {volume} {80}},\ \bibinfo {pages} {064419}
  (\bibinfo {year} {2009})}\BibitemShut {NoStop}%
\bibitem [{\citenamefont {Tocchio}\ \emph {et~al.}(2013)\citenamefont
  {Tocchio}, \citenamefont {Feldner}, \citenamefont {Becca}, \citenamefont
  {Valent\'{\i}},\ and\ \citenamefont {Gros}}]{Tocchio2013}%
  \BibitemOpen
  \bibfield  {author} {\bibinfo {author} {\bibfnamefont {L.~F.}\ \bibnamefont
  {Tocchio}}, \bibinfo {author} {\bibfnamefont {H.}~\bibnamefont {Feldner}},
  \bibinfo {author} {\bibfnamefont {F.}~\bibnamefont {Becca}}, \bibinfo
  {author} {\bibfnamefont {R.}~\bibnamefont {Valent\'{\i}}}, \ and\ \bibinfo
  {author} {\bibfnamefont {C.}~\bibnamefont {Gros}},\ }\href {\doibase
  10.1103/PhysRevB.87.035143} {\bibfield  {journal} {\bibinfo  {journal} {Phys.
  Rev. B}\ }\textbf {\bibinfo {volume} {87}},\ \bibinfo {pages} {035143}
  (\bibinfo {year} {2013})}\BibitemShut {NoStop}%
\bibitem [{\citenamefont {Leung}\ and\ \citenamefont
  {Runge}(1993)}]{Leung1993}%
  \BibitemOpen
  \bibfield  {author} {\bibinfo {author} {\bibfnamefont {P.~W.}\ \bibnamefont
  {Leung}}\ and\ \bibinfo {author} {\bibfnamefont {K.~J.}\ \bibnamefont
  {Runge}},\ }\href {\doibase 10.1103/PhysRevB.47.5861} {\bibfield  {journal}
  {\bibinfo  {journal} {Phys. Rev. B}\ }\textbf {\bibinfo {volume} {47}},\
  \bibinfo {pages} {5861} (\bibinfo {year} {1993})}\BibitemShut {NoStop}%
\bibitem [{\citenamefont {Bernu}\ \emph {et~al.}(1994)\citenamefont {Bernu},
  \citenamefont {Lecheminant}, \citenamefont {Lhuillier},\ and\ \citenamefont
  {Pierre}}]{Bernu1994}%
  \BibitemOpen
  \bibfield  {author} {\bibinfo {author} {\bibfnamefont {B.}~\bibnamefont
  {Bernu}}, \bibinfo {author} {\bibfnamefont {P.}~\bibnamefont {Lecheminant}},
  \bibinfo {author} {\bibfnamefont {C.}~\bibnamefont {Lhuillier}}, \ and\
  \bibinfo {author} {\bibfnamefont {L.}~\bibnamefont {Pierre}},\ }\href
  {\doibase 10.1103/PhysRevB.50.10048} {\bibfield  {journal} {\bibinfo
  {journal} {Phys. Rev. B}\ }\textbf {\bibinfo {volume} {50}},\ \bibinfo
  {pages} {10048} (\bibinfo {year} {1994})}\BibitemShut {NoStop}%
\bibitem [{\citenamefont {Heidarian}\ \emph {et~al.}(2009)\citenamefont
  {Heidarian}, \citenamefont {Sorella},\ and\ \citenamefont
  {Becca}}]{Heidarian2009}%
  \BibitemOpen
  \bibfield  {author} {\bibinfo {author} {\bibfnamefont {D.}~\bibnamefont
  {Heidarian}}, \bibinfo {author} {\bibfnamefont {S.}~\bibnamefont {Sorella}},
  \ and\ \bibinfo {author} {\bibfnamefont {F.}~\bibnamefont {Becca}},\ }\href
  {\doibase 10.1103/PhysRevB.80.012404} {\bibfield  {journal} {\bibinfo
  {journal} {Phys. Rev. B}\ }\textbf {\bibinfo {volume} {80}},\ \bibinfo
  {pages} {012404} (\bibinfo {year} {2009})}\BibitemShut {NoStop}%
\bibitem [{\citenamefont {Thesberg}\ and\ \citenamefont
  {S\o{}rensen}(2014)}]{Thesberg2014}%
  \BibitemOpen
  \bibfield  {author} {\bibinfo {author} {\bibfnamefont {M.}~\bibnamefont
  {Thesberg}}\ and\ \bibinfo {author} {\bibfnamefont {E.~S.}\ \bibnamefont
  {S\o{}rensen}},\ }\href {\doibase 10.1103/PhysRevB.90.115117} {\bibfield
  {journal} {\bibinfo  {journal} {Phys. Rev. B}\ }\textbf {\bibinfo {volume}
  {90}},\ \bibinfo {pages} {115117} (\bibinfo {year} {2014})}\BibitemShut
  {NoStop}%
\bibitem [{\citenamefont {Mendes-Santos}\ \emph {et~al.}(2015)\citenamefont
  {Mendes-Santos}, \citenamefont {Paiva},\ and\ \citenamefont {dos
  Santos}}]{Mendes-Santos2014}%
  \BibitemOpen
  \bibfield  {author} {\bibinfo {author} {\bibfnamefont {T.}~\bibnamefont
  {Mendes-Santos}}, \bibinfo {author} {\bibfnamefont {T.}~\bibnamefont
  {Paiva}}, \ and\ \bibinfo {author} {\bibfnamefont {R.~R.}\ \bibnamefont {dos
  Santos}},\ }\href {\doibase 10.1103/PhysRevA.91.023632} {\bibfield  {journal}
  {\bibinfo  {journal} {Phys. Rev. A}\ }\textbf {\bibinfo {volume} {91}},\
  \bibinfo {pages} {023632} (\bibinfo {year} {2015})}\BibitemShut {NoStop}%
\end{thebibliography}%
\end{document}